\documentclass [smallextended] {article}
\usepackage{amsmath}
\usepackage{physics}
\usepackage{times}
\usepackage{ifpdf}
\usepackage[english]{babel}
\usepackage{braket}
\usepackage{listings}
\usepackage{xcolor}
\usepackage{algorithm}
\usepackage[noend]{algpseudocode}
\usepackage{hyperref}
\usepackage{graphicx}
\usepackage{mathrsfs}

\usepackage[square,numbers]{natbib}
\definecolor{codegreen}{rgb}{0,0.6,0}
\definecolor{codegray}{rgb}{0.5,0.5,0.5}
\definecolor{codepurple}{rgb}{0.58,0,0.82}
\definecolor{backcolour}{rgb}{0.95,0.95,0.92}
 
\lstdefinestyle{mystyle}{
    backgroundcolor=\color{backcolour},   
    commentstyle=\color{codegreen},
    keywordstyle=\color{black},
    numberstyle=\tiny\color{codegray},
    stringstyle=\color{codepurple},
    basicstyle=\ttfamily\footnotesize,
    breakatwhitespace=false,         
    breaklines=true,                 
    captionpos=b,                    
    keepspaces=true,                 
    numbers=none,                    
    numbersep=5pt,                  
    showspaces=false,                
    showstringspaces=false,
    showtabs=false,                  
    tabsize=2
}
 
\newcommand{\Delete}[1]{}
\newcommand{\hidecode}[1]{}

\title{A Quantum Vocal Theory of Sound}
\author{Davide Rocchesso\footnote{davide.rocchesso@unipa.it}\,\, and Maria Mannone \\ Department of Mathematics and Computer Science, \\ University of Palermo, Italy}
\begin{document}

\maketitle

\begin{abstract}
 Concepts and formalism from acoustics are often used to exemplify quantum mechanics. Conversely, quantum mechanics could be used to achieve a new perspective on acoustics, as shown by Gabor studies. Here, we focus in particular on the study of human voice, considered as a probe to investigate the world of sounds. We present a theoretical framework that is based on {\em observables} of vocal production, and on some {\em measurement apparati} that can be used both for analysis and synthesis. In analogy to the description of spin states of a particle, the quantum-mechanical formalism is used to describe the relations between the fundamental states associated with phonetic labels such as phonation, turbulence, and supraglottal myoelastic vibrations. The intermingling of these states, and their temporal evolution, can still be interpreted in the Fourier/Gabor plane, and effective extractors can be implemented. The bases for a Quantum Vocal Theory of Sound, with implications in sound analysis and design, are presented. 
\end{abstract}

\section{Introduction}
What are the fundamental elements of sound? What is the most meaningful framework for analyzing existing sonic realities and for expressing new sound concepts? These are long standing questions in sound physics, perception, and creation. In his analytical theory of heat~\cite{fourier1822}, Joseph Fourier laid the basis for analyzing functions of one variable in terms of sinusoidal components, and explicitly wrote that ``\dots if the order which is established in these phenomena could be grasped by our senses, it would produce in us an impression comparable to the sensation of musical sounds''. 

Hermann von Helmholtz took Fourier's suggestion seriously and proceeded to analyze all vibratory phenomena as additions of sinusoidal vibrations~\cite{helmoltz1870}. Although he admitted that ``we can conceive a whole to be split into parts in very different and arbitrary ways'', it was the observation that the ear somehow reflects Fourier analysis, and can be described as a bank of sympathetic resonators, that led him to state that ``the existence of partial tones [\dots] acquire a meaning in nature''.

In the twentieth century, despite the Fourier transform being the key to describe sampling and signal reconstruction from samples~\cite{shannon1949}, skepticism arose among physicists such as Norbert Wiener and Dennis Gabor about considering Fourier analysis as the best representation for music~\cite{roads2001}. In 1947, in a famous paper published in Nature~\cite{gabor}, Gabor embraced the mathematics of quantum theory to shed light on subjective acoustics, thus laying the basis for sound analysis and synthesis based on acoustical quanta, or grains, or wavelets. 

The Fourier and Gabor frameworks for time-frequency, or time-scale, representation of sound are widely used in the analysis and synthesis of sonic phenomena. For example, the auditory time and frequency acuities have been bounded in terms of the uncertainty principle, although the theoretical limit has been shown to beaten by human audition~\cite{OppenheimMagnasco2013}. As another example, cochlear filters are designed so that their time-frequency behavior matches human performance, and are used to simulate or replace human hearing~\cite{lyon}. 

Still, when we are imagining a sound, or describing it to peers, we do not use the Fourier formalism, but we rather refer to the hypothetical sources and to their characteristics~\cite{gaver}, or we use our voice to mimic some salient sound features, thus overcoming the limitations of language~\cite{lemaitre14}. We argue, therefore, that a description of sound that exploits the basic mechanisms of voice production would be more readily understandable and manipulable than any decomposition based on framed sines or on chirps.

In this contribution, we propose a phonetic approach to describe sound at large. A coarse articulatory description can indeed be applied to any sound, and it will provide the basis for attempting a vocal imitation, which makes embodied sound perception concrete and audible. In presence of concurrent sources, the high-level phonetic descriptors are superimposed and temporally varying, and their evolution is governed by context and attention. Apparently, our hearing system acts as a sort of destructive measurement apparatus which continuously collapses superpositions of phonetic states into streams~\cite{bregman1994auditory}, whose evolution we can single out and follow, with the possibility of jumping from one stream to another as a result of hidden or apparent forces. 

Superposition and evolution of states, together with the concepts of measurement collapse and force fields, are among the cornerstones of quantum theory, and this is the observation that led us to attempt a description of sonic phenomena within a quantum framework. Hopefully, some phenomena that are normally described through sets of rules and gestalt principles (e.g., auditory continuity or temporal displacement) may naturally emerge from such a quantum-inspired description, similarly to how quantum cognition has been able to address behaviors that are difficult to derive within classical frameworks~\cite{yearsley2016}.
The apparent incompatibility of properties that are being judged, or of forms that are being perceived, implies some vagueness in the mental states and in their time evolution, which is difficult to model classically but is intrinsic in quantum modeling~\cite{yearsleyPothos2014}. This is particularly evident in bivalued judgements and in bistable percepts, that can be modeled as a two-state quantum-mechanical system, or qubit.
We intend to apply such a quantum-theoretical model, which is constructed in analogy with spins in a time-varying magnetic field, to auditory scenes made of overlapping auditory objects, described in phonetic terms. In the context of auditory scene analysis, we introduce the quantum-theoretical concepts of superposition, time evolution, and measurement (or foreground separation). We show how  this framework can be useful to describe and reproduce some auditory-streaming phenomena, with possible applications in source separation and audio effects. 

Section~\ref{background} provides a short background on prior research on the two main axes that cross in this work: Research in sound objecthood, with special emphasis on the voice as an embodied representation of sound; Quantum frameworks that have been proposed for sound and image processing, music, and perception. Section~\ref{sketchQVTS} gives the motivation and a compact overview of the proposed Quantum Vocal Theory of Sound. The long section~\ref{sec:formalism} recalls the basic mathematical formalism and some key concepts of quantum theory, and it shows how these tools and concepts can be recast in audio terms.  Section~\ref{examples} shows how quantum evolution can inspire algorithms for auditory object streaming and separation, thus pointing to possible applications in computational auditory scene analysis and audio effects.

\section{Background}\label{background}
\subsection{Voice as Embodied Sound}
Many researchers, in science, art, and philosophy, have been facing the problem of how to approach sound and its representations \cite{dePoli,roden}. Should we represent sounds as they appear to the senses, by manipulating their proximal characteristics? Or should we rather look at potential sources, at physical systems that produce sound as a side effect of distal interactions? In this research path we assume that our body can help establish bridges between distal (source-related) and proximal (sensory-related) representations, and we look at research findings in perception, production, and articulation of sounds~\cite{leman,valle2015}. Our approach to sound \cite{delleMonache2018,rocchesso2019} seeks to exploit knowledge in these areas, especially referring to human voice production as a form of embodied representation of sound.

When considering what people hear from the environment, it emerges that sounds are mostly perceived as belonging to categories of the physical world \cite{gaver}. Research in sound perception has shown that listeners spontaneously create categories such as solid, electrical, gas, and liquid sounds, even though the sounds within these categories may be acoustically different \cite{houix}. However, when the task is to separate, distinguish, count, or compose sounds, the attention shifts from {\em sounding objects} to {\em auditory objects}~\cite{kubovy} represented in the time-frequency plane, or to {\em auditory images}, which are movie-like temporal representations resembling the signals projected by the ear up to the auditory cortex~\cite{lyon}. Tonal components, noise, and transients can be extracted from auditory objects with Fourier-based techniques \cite{bonada2011spectral,verma,fug2016}. Low-frequency periodic phenomena are also perceptually very relevant and often come as trains of transients. The most prominent elements of the proximal signal may be selected by simplification and inversion of time-frequency representations. These auditory sketches~\cite{isnard} have been used to test the recognizability of imitations~\cite{lemaitre2016plos}.

When discussing spaces for sound representation it is also important to recall the notion of {\it sound object}, often associated to Schaeffer's theory of listening and typo-morphological spaces, which support a phenomenological description of sound and can be reported to the time-frequency plane~\cite{valle2015}. For example, the concept of mass is a generalization of the notion of pitch that comprises both site (on the frequency axis) and caliber (or degree of occupation of the frequency axis).

Vocal imitations can be more effective than verbalizations at representing and communicating sounds when these are difficult to describe with words \cite{lemaitre14}. This indicates that vocal imitations can be a useful tool for investigating sound perception, and shows that the voice is instrumental to embodied sound cognition. 
Vocal imitations act similarly to visual sketches: they catch and emphasize some essential elements of the original (visual) objects allowing their identification. 
At a more fundamental level, research on non-speech vocalization is affecting the theories of language evolution \cite{perlman}, as it seems plausible that humans could have used iconic vocalizations to communicate with a large semantic spectrum, prior to the establishment of full-blown spoken languages. Experiments and sound design exercises~\cite{delleMonache2018} show that agreement in production corresponds to agreement in meaning interpretation, thus showing the effectiveness of teamwork in embodied sound creation. Converging evidences from behavioral and brain imaging studies give a firm basis to hypothesize a shared representation of sound in terms of motor (vocal) primitives~\cite{wallmark}. Historically, such convergence was envisioned over a century ago by the Italian Futurists: On one side, the composer Luigi Russolo developed an organology of everyday sounds and devised mechanical synthesizers for these ``noises''~\cite{russolo1916}; On the other side, the poet Filippo Tommaso Marinetti devised a way to transcend language to bring everyday sounds to poetry, through imitations and onomatopeia~\cite{marinetti1914}.

Some phoneticians have turned their attention to non-speech voice production, trying to identify the most relevant phonetic components that are found in vocal imitations~\cite{helgason}. They identified the broad categories of phonation (i.e., quasi periodic oscillations due to vocal fold vibrations), turbulence, supraglottal myoelastic vibrations, and clicks, which can be extracted automatically from audio with time-frequency analysis and supervised~\cite{friberg2018} or unsupervised~\cite{marchetto} machine learning. These categories can be made to correspond to categories of sounds as they are perceived~\cite{lemaitre16}, and as they are produced in the physical world. Indeed, it has been argued that human utterances somehow mimic ``nature's phonemes''~\cite{changizi}, and neurophysiological studies have shown that the cortical area of the superior temporal gyrus actually encodes abstract phonetic features~\cite{mesgarani2014}.

\subsection{Quantum Frameworks}
It was Dennis Gabor \cite{gabor} who first adopted the mathematics of quantum mechanics to explain acoustic phenomena. In particular, he used operator methods to derive the time-frequency uncertainty relation and the (Gabor) function that satisfies minimal uncertainty. Time-scale representations \cite{deSena} are more suitable to explain the perceptual decoupling of pitch and timbre, and operator methods can be used as well to derive the gammachirp function, which minimizes uncertainty in the time-scale domain \cite{irino}. Research in human and machine hearing \cite{lyon} have been based on banks of elementary (filter) functions and these systems are at the core of many successful applications in the audio domain.

Despite its deep roots in the physics of the twentieth century, the sound field has not yet embraced the quantum signal processing framework \cite{eldar} to seek practical solutions to sound scene representation, separation and analysis, although some theoretical proposals to encode, store, and process audio using quantum circuitry have been advanced~\cite{wang2016,yan2018}. 
On the other hand, some common observed properties of human cognition and quantum mechanics (superposition, non-classical probability) have given universal value to the quantum-theoretic formalism to explain cognitive acts~\cite{yearsley2016}, including actions of human creation, such as music. The explanatory power of a quantum approach to music cognition has been demonstrated to describe tonal attraction phenomena in terms of metaphorical forces~\cite{beimGraben19,blutnerGraben20}. The theory of open quantum systems has been applied to music to describe the memory properties (non-Markovianity) of different scores \cite{mannone}. The time-dependent Schr\"{o}dinger equation for a single non-relativistic particle has been used as a model for sound and music composition. Some examples include the creation of grain clouds like orbitals~\cite{fischman}, the sonification of controlled quantum dynamics~\cite{konto},
and compositions for an ensemble of atoms~\cite{sturm}.
 It has even been claimed that the interplay between musical ideas and extra-musical meanings can be naturally represented in the framework of quantum semantics, where extra-musical meanings can be treated within a theory of vague possible worlds~\cite{dallaChiara2015}. 

Some theoretical physicists have looked at the sensory processes driving human and animal perception, trying to understand if they are classical or quantum. As far as visual perception is concerned, Ghirardi proposed an experiment to verify if the perceptive apparatus can induce the suppression of a physically-established superposition of states~\cite{GHIRARDI1999}. 
In application-oriented image processing, on the other hand, it has been shown how the quantum framework can be effective to solve problems such as segmentation. For example, the separation of figures from background can be obtained by evolving a solution of the time-dependent Schr\"{o}dinger equation~\cite{youssry}, or by discretizing the time-independent Schr\"{o}dinger equation~\cite{aytekin}. An approach to signal manipulation based on the postulates of quantum mechanics can also potentially lead to a computational advantage when using Quantum Processing Units. Results in this direction are being reported for optimization problems~\cite{okada2019}.

In this work, we consider auditory phenomena and look at quantum theory for a possible process model, that somehow mirrors the way humans extract and follow auditory objects from audio mixtures. Such a process model, that exploits our embodied knowledge of sound via vocal production, does not assume any underlying information processing model for the brain. This standpoint and disclaimer is commonly assumed in quantum cognition~\cite{yearsley2016} and readily adopted here.

\section{Sketch of a Quantum Vocal Theory of Sound}
\label{sketchQVTS}
In the proposed research path, sound is treated as a superposition of states, and the voice-based components (phonation, turbulence, supraglottal myoelastic vibrations) are considered as observables to be represented as operators. The extractors of the fundamental components, i.e., the measurement apparati, are implemented as signal-processing modules that are available both for analysis and, as control knobs, for synthesis.
The baseline is found in the results of the SkAT-VG project~\cite{rocchesso2015,lemaitre14,lemaitre16,lemaitre2016plos,delleMonache2018,friberg2018}, which showed that vocal imitations are optimized representations of referent sounds, that emphasize those features that are important for identification. A large collection of audiovisual recordings of vocal and gestural imitations\footnote{\href{https://www.ircam.fr/projects/blog/multimodal-database-of-vocal-and-gestural-imitations-elicited-by-sounds/}{https://www.ircam.fr/projects/blog/multimodal-database-of-vocal-and-gestural-imitations-elicited-by-sounds/}} offers the opportunity to further enquire how people perceive, represent, and communicate about sounds.

A first assumption underlying this research approach, largely justified by prior art and experiences, is that articulatory primitives used to describe vocal utterances are effective as high-level descriptors of sound in general. This assumption leads naturally to an embodied approach to sound representation, analysis, and synthesis. 

A second assumption is that the mathematics of quantum mechanics, relying on linear operators in Hilbert spaces, offers a formalism that is suitable to describe the objects composing auditory scenes and their evolution in time. The latter assumption is more adventurous, as this path has not been taken in audio signal processing yet. However, the results coming from neighboring fields (music cognition, image processing) encourage us to explore this direction, and to aim at introducing new techniques for sound analysis, synthesis, and transformation.

An embryonic theory of sound based on the postulates of quantum
mechanics, and using high-level vocal descriptors of sound, can be
sketched as follows. Let $\overline{\sigma}$ be a vector operator that provides
information about the phonetic elements along a specific {direction}
of measurement. Phonation, for example, may be represented by
$\sigma_z$, with eigenstates representing a upper and a lower
pitch. Similarly, the turbulence component may be represented by
$\sigma_x$, with eigenstates representing turbulence of two different spectral
distributions. A measurement of turbulence prepares the {\em system} in one of two eigenstates for operator $\sigma_x$, and a successive
measurement of phonation would find a superposition and get equal
probabilities for the two eigenstates of $\sigma_z$. The two operators
$\sigma_z$ and $\sigma_x$ may also be made to correspond to the two
components of the classic sines + noise model used in audio signal
processing. If we add transients/clicks as a third measurement
direction (as in the sines + noise + transients model~\cite{verma}) we
can claim that there is no sound state for which the expectation value
of the three components is zero: a sort of spin polarization principle
as found in quantum mechanics. The evolution of state vectors in time
is unitary, and regulated by a time-dependent Schr\"{o}dinger
equation, with a suitably chosen Hamiltonian. The eigenvectors of the
Hamiltonian allow to expand any state vector in that basis, and to
compute the time evolution of such expansion. A pair of components can
be simultaneously measured only if they commute. If they don't, an
uncertainty principle can be derived, as it was done for
time-frequency and time-scale representations~\cite{gabor,irino}. The theory can be
extended to cover multiple uncertain sources, and the resulting mixed states can
be described via density matrices, whose time evolution can also be
computed if a Hamiltonian operator is properly defined. In the following, we formally lay down this Quantum Vocal Theory of Sound.
\\
\\
\section{The phon formalism}\label{sec:formalism}
Consider a 3d space with the orthogonal axes
\begin{description}
\item[z]: phonation, with different pitches;
\item[x]: turbulence, with different brightnesses;
\item[y]: myoelasticity, slow pulsations with different tempos.
\end{description}
The labels attributed to the axes correspond to the three main
articulatory/phonatory categories that are used by phoneticians to
annotate vocal imitations of everyday sounds~\cite{helgason}. They are
a simplification of the more phonetically-correct labels ``vocal fold
phonation,'' ``turbulence,'' and ``supraglottal myoelastic vibration''~\cite{friberg2018}. 

The phon operator $\overline{\sigma}$ is a 3-vector operator that
provides information about the phonetic component in a specific
direction of the 3d phonetic space, i.e., along a specific combination
of phonation, turbulence, and myoelasticity.

In this section we present the phon formalism, obtained by direct
analogy with the single spin, as presented in accessible presentations
of quantum mechanics~\cite{susskind2014quantum}. We use standard Dirac notation and adopt the quantum-theoretical concepts of measurement, preparation, pure and mixed states, uncertainty, and time evolution~\cite{cariolaro2015quantum}.

\subsection{Measurement along z}
A measurement along the z axis is performed according to the
quantum-mechanics principles:
\begin{enumerate}
\item Each component of $\overline{\sigma}$  is represented by a linear operator;
\item The eigenvectors of $ \sigma_z $ are $\ket{u}$ and $\ket{d}$,
  corresponding to pitch-up and pitch-down, with eigenvalues $+1$ and
  $-1$, respectively:
  \begin{enumerate}
    \item $ \sigma_z \ket{u} = \ket{u}$
    \item $ \sigma_z \ket{d} = - \ket{d}$
  \end{enumerate}
\item The eigenstates of operator  $ \sigma_z $, $ \ket{u} $ and $ \ket{d} $, are
  orthogonal:  $\braket{u|d} = 0 $;
\end{enumerate}
The eigenstates can be represented as column vectors 
$$\ket{u}
= \begin{bmatrix}1\\0\end{bmatrix}, \, \ket{d}
  = \begin{bmatrix}0\\1\end{bmatrix},$$ and the operator $
    \sigma_z $ as a square $2 \times 2$ matrix. Due to principle 2,
    we have
    \begin{equation}
      \sigma_z = \begin{bmatrix} 1 &  0 \\ 0 & -1 \end{bmatrix}.
      \label{pauli1}
    \end{equation}
    
\subsection{Preparation along x}
The eigenstates of the operator $\sigma_x$ are $ \ket{r} $ and $
\ket{l} $, corresponding to turbulences having different spectral
distributions, one with the rightmost (or highest-frequency) centroid and the other with the
leftmost centroid. The respective eigenvalues are $+1$ and $-1$, so that
\renewcommand{\labelenumi}{(\alph{enumi})}
  \begin{enumerate}
    \item $ \sigma_x \ket{r} = \ket{r}$
    \item $ \sigma_x \ket{l} = - \ket{l}$ .
  \end{enumerate}
If the phon is prepared $\ket{r}$ (turbulent) and then the measurement
apparatus is set to measure $\sigma_z$, there will be equal
probabilities for $\ket{u}$ or $\ket{d}$ phonation as an
outcome. Essentially, we are measuring what kind of phonation is in a
pure turbulent state. This measurement property is satisfied if
\begin{equation}
  \ket{r} = \frac{1}{\sqrt{2}} \ket{u} +  \frac{1}{\sqrt{2}} \ket{d}.
\label{rtoud}
\end{equation}
Likewise, if the phon is prepared $\ket{l}$  and then the measurement
apparatus is set to measure $\sigma_z$, there will be equal
probabilities for $\ket{u}$ or $\ket{d}$ phonation as an
outcome.  This measurement property is satisfied if
\begin{equation}
  \ket{l} = \frac{1}{\sqrt{2}} \ket{u} -  \frac{1}{\sqrt{2}} \ket{d},
\end{equation}
which is orthogonal to the linear combination~(\ref{rtoud}).
In vector form, we have
$$\ket{r}
= \begin{bmatrix}\frac{1}{\sqrt{2}}\\\frac{1}{\sqrt{2}}\end{bmatrix},\,\, \ket{l}
  = \begin{bmatrix}\frac{1}{\sqrt{2}}\\-\frac{1}{\sqrt{2}}\end{bmatrix},  \mbox{ and }
  $$
  \begin{equation}
    \sigma_x = \begin{bmatrix} 0 &  1 \\ 1 & 0 \end{bmatrix}.
    \label{pauli2}
  \end{equation}

  In fact, any state $\ket{A}$ can be expressed as
  \begin{equation}
    \ket{A} = \alpha_u \ket{u} + \alpha_d \ket{d},
    \label{anystate}
  \end{equation}
  where $\alpha_u = \braket{u|A}$, and $\alpha_d =
  \braket{d|A}$. Being the system in state $\ket{A}$, the probability
  to measure pitch-up is
  \begin{equation}
    p_u = \braket{A|u}\braket{u|A} = {\alpha_u}^*\alpha_u
  \end{equation}
  and, similarly, the probability to measure pitch-down is $p_d =
  \braket{A|d}\braket{d|A} = {\alpha_d}^*\alpha_d$ (Born rule).

\subsection{Preparation along y}
The eigenstates of the operator $\sigma_y$ are $ \ket{f} $ and $
\ket{s} $, corresponding to slow myoelastic pulsations, one faster and one slower\footnote{In describing the spin eigenstates, the symbols $\ket{i}$ and $\ket{o}$ are often used, to denote the in--out direction.}, with eigenvalues $+1$ and $-1$, so that
  \begin{enumerate}
    \item $ \sigma_y \ket{f} = \ket{f}$
    \item $ \sigma_y \ket{s} = - \ket{s}$ .
  \end{enumerate}
If the phon is prepared $\ket{f}$ (pulsating) and then the measurement
apparatus is set to measure $\sigma_z$, there will be equal
probabilities for $\ket{u}$ or $\ket{d}$ phonation as an
outcome. Essentially, we are measuring what kind of phonation is in a
myoelastic pulsations. This measurement property is satisfied if
\begin{equation}
  \ket{f} = \frac{1}{\sqrt{2}} \ket{u} +  \frac{i}{\sqrt{2}} \ket{d},
\label{itoud}
\end{equation}
where $i$ is the imaginary unit.

Likewise, if the phon is prepared $\ket{s}$, we can express this state as
\begin{equation}
  \ket{s} = \frac{1}{\sqrt{2}} \ket{u} -  \frac{i}{\sqrt{2}} \ket{d},
\end{equation}
which is orthogonal to the linear combination~(\ref{itoud}).
In vector form, we have
$$\ket{f}
= \begin{bmatrix}\frac{1}{\sqrt{2}}\\\frac{i}{\sqrt{2}}\end{bmatrix},\,\, \ket{s}
  = \begin{bmatrix}\frac{1}{\sqrt{2}}\\-\frac{i}{\sqrt{2}}\end{bmatrix}, \mbox{ and }
  $$
  %\newline and 
  \begin{equation}
    \sigma_y = \begin{bmatrix} 0 &  -i \\ i & 0 \end{bmatrix}.
    \label{pauli3}
  \end{equation}
The matrices~(\ref{pauli1}), (\ref{pauli2}), and (\ref{pauli3}) are called the Pauli matrices and, together with the identitity matrix, these are the quaternions.

\subsection{Measurement along an arbitrary direction}
Orienting the measurement apparatus along an arbitrary direction
$\overline{n} = \left[n_x, n_y, n_z\right]'$ means taking a weighted mixture of quaternions:
\begin{multline}
  \sigma_n = \overline{\sigma} \cdot \overline{n} = \sigma_x n_x + \sigma_y n_y + \sigma_z n_z  = \begin{bmatrix} n_z &  n_x - i n_y \\ n_x + i n_y & -n_z \end{bmatrix}. 
\end{multline}

\subsubsection{Example: Harmonic plus Noise model}
A measurement performed by means of a Harmonic plus Noise model~\cite{bonada2011spectral}
would lie in the phonation-turbulence plane ($n_z = \cos\theta, n_x = \sin\theta, n_y = 0$), so that
\begin{equation}
  \sigma_n = \begin{bmatrix} \cos \theta & \sin \theta \\ \sin \theta & -\cos \theta \end{bmatrix}
\end{equation}
The eigenstate for eigenvalue  $+1$ is \begin{align} \ket{\lambda_1}
= \left[ \cos \theta / 2,   \sin \theta / 2 \right]',\end{align}
the eigenstate for eigenvalue $-1$ is \begin{align}\ket{\lambda_{-1}}
= \left[ - \sin \theta / 2, \cos \theta / 2 \right]',\end{align}
and the two are orthogonal.
Suppose we prepare the phon to pitch-up $\ket{u}$. If we rotate the measurement system along $\overline{n}$, the probability to measure $+1$ is (by Born rule)
\begin{equation}
  p(+1) = \left|\braket{u|\lambda_1}\right|^2 = \cos^2 \theta/2,
\end{equation}
and  the probability to measure $-1$ is 
\begin{equation}
  p(-1) = \left|\braket{u|\lambda_{-1}}\right|^2 = \sin^2 \theta/2.
\end{equation}
The expectation value of measurement is therefore
    \begin{multline}
    \braket{\sigma_n} = \sum_j \lambda_j p(\lambda_j) = (+1) \cos^2 \theta/2 + (-1) \sin^2 \theta/2 = \cos \theta .
  \end{multline}

\subsubsection{Rotate to measure}
What does it mean to rotate a measurement apparatus to measure a
property? Assume we have a machine that separates harmonics from noise
from (trains of) transients, and that can discriminate between two
different pitches, noise distributions, and tempos. Essentially, the
machine receives a sound and returns three numbers $\{\rm{ph},
\rm{tu}, \rm{my}\} \in [-1, 1]$. If $\rm{ph} > 0$ the result will be
$\ket{u}$, and if $\rm{ph} < 0$ the result will be $\ket{d}$. If
$\rm{tu} > 0$, the result will be $\ket{r}$, and if $\rm{tu} < 0$ the
result will be $\ket{l}$. If $\rm{my} > 0$, the result will be
$\ket{f}$ and if $\rm{my} < 0$, the result will be $\ket{s}$. These
three outputs correspond to rotating the measurement apparatus along
each of the main axes. Rotating it along an arbitrary direction means
taking a weighted mixture of the three outcomes.

\begin{figure}
 \centerline{{
     \includegraphics[width=\columnwidth]{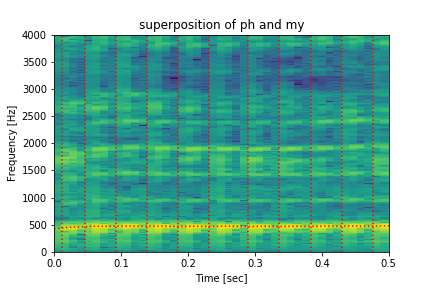}}}
 \caption{Spectrogram of a vocal sound which is a superposition of phonation and supgraglottal myoelastic vibration. A salient pitch (horizontal dotted line) as well as quasi-regular train of pulses (vertical dotted lines) are automatically extracted.}
 \label{superposition_ph_my} 
\end{figure}

For example, consider the vocal fragment\footnote{It is one of the example vocal sounds considered in~\cite{rocchesso2016}, and taken from~\cite{newman2004}.} whose spectrogram is represented in fig.~\ref{superposition_ph_my}. An extractor of pitch salience can be used to measure phonation, and an extractor of onsets can be used to measure slow myoelastic pulsation. Such two feature extractors, as found in the Essentia library~\cite{bogdanov2013essentia}, have been applied to highlight the phonation (horizontal dotted line) and myoelastic (vertical dotted lines) components in the spectrogram of fig.~\ref{superposition_ph_my}. {In the $z - y$ plane, there would be a measurement orientation and a measurement operator that admits such sound as an eigenvector.}

\subsection{Pure and Mixed states}
According to the first postulate of quantum
mechanics~\cite{cariolaro2015quantum}, at each time instant the system
is completely specified by a state $\ket{\psi}$ such that
$\braket{\psi | \psi} = 1$.  If the state is known with certainty it
is called a pure state. All the phon states described so far are pure
states. More generally, a state can be known probabilistically as one
of a set of $\ket{\psi_i}$ with a given probability
distribution. States of such kind are called mixed states. The density
operator represents both pure and mixed states, and it is defined as
\begin{equation}
  \rho = \sum_j p_j \ket{\psi_j} \bra{\psi_j},
\end{equation}
where $p_j$ is probability for state $\ket{\psi_j}$.

For a pure state, it is simply $\rho = \ket{\psi} \bra{\psi}$, and the
trace of the square of such matrix is $Tr[\rho^2] = 1$.
For a {mixed state}, it is always the case that $Tr[\rho^2]
< 1$.

\subsubsection{Example}
Let state $\ket{u} $ with probability $\frac{1}{3}$ and state
$\ket{d}$ with probability $\frac{2}{3}$. The density matrix is
\begin{equation}\label{exampleRho}
\rho = \frac{1}{3} \ket{u} \bra{u} + \frac{2}{3} \ket{d} \bra{d}
= \begin{bmatrix} \frac{1}{3} & 0 \\ 0 & \frac{2}{3} \end{bmatrix},
\end{equation}
and the trace of its square is
$$Tr[\rho^2] = \frac{5}{9} < 1. $$

The interest of the density operator is given by its generalization power. It is an essential generalization in quantum mechanics and, as such, it is relevant for a quantum vocal theory of sound. From an experimental point of view, it introduces a degree of conceptual flexibility which may come useful in synthesis and composition of auditory scenes. In particular, the audio concept of mixing can be made to correspond with manipulation of mixed states.

\subsection{Uncertainty}
If we measure two observables ${\bf L}$ and ${\bf M}$ (in a single experiment)
simultaneously, quantum mechanics prescribes that the system is left
in a simultaneous eigenvector of the observables only if ${\bf L}$ and
${\bf M}$ commute, i.e., if their commutator $\left[ {\bf L, M} \right] = {\bf LM - ML}$ is null.  Measurement
operators along different axes do not commute. For example, $\left[
  \sigma_x, \sigma_y \right] = 2 i \sigma_z$, and therefore phonation
and turbulence can not be simultaneously measured with certainty.

The uncertainty principle, based on Cauchy-Schwarz inequality in
complex vector spaces, prescribes that the product of the two
uncertainties is at least as large as half the magnitude of the
commutator:
\begin{equation} \label{uncertaintyPrinciple}
  \Delta {\bf L} \Delta {\bf M} \geq \frac{1}{2} \left| \braket{\psi | \left[{\bf L, M}\right] | \psi} \right|
\end{equation}

If ${\bf L} = \mathscr{T} = t$ is the time operator and ${\bf M} = \mathscr{W} = -i \frac{d}{dt}$ is the
frequency operator, and these are applied to the complex oscillator $A
e^{i \omega t}$, the time-frequency uncertainty principle results, and
uncertainty is minimized by the Gabor function. Starting from the
scale operator, the gammachirp function can be derived~\cite{irino}.

\subsection{Time evolution}
Another postulate of quantum mechanics~\cite{cariolaro2015quantum}
states that the evolution of state vectors in time
\begin{equation} \label{unitaryTransformation}
  \ket{\psi(t)} = {\bf U}(t_0, t) \ket{\psi(t_0)}, t > t_0,
\end{equation}
is governed by the operator ${\bf U}$, which is unitary (i.e., ${\bf
  U}^\dagger {\bf U} = {\bf I}$) and depends only on $t_0$ and $t$.
Taken a small time increment $\epsilon$, continuity of the
time-development operator gives it the form
\begin{equation}  \label{eq:Uepsilon}
  {\bf U}(\epsilon) = {\bf I} - i \epsilon {\bf H}, 
\end{equation}
with ${\bf H}$ being the quantum Hamiltonian (Hermitian)
operator. ${\bf H}$ is an observable and its eigenvalues are the values
that would result from measuring the energy of a quantum
system. From~(\ref{eq:Uepsilon}) it turns out that a state vector
changes in time according to the time-dependent Schr\"odinger
equation\footnote{We do not need physical dimensional consistency
  here, so we drop Planck's constant.}
\begin{equation}    \label{eq:timedepSchr}
  \frac{\partial
  \ket{\psi(t)}}{\partial t} = - i {\bf H}(t) \ket{\psi(t)}. 
\end{equation}
Any observable ${\bf L}$ has an expectation value $\braket{\bf L}$ that
evolves according to
\begin{equation} \label{eq:Lcomm}
  \frac{\partial \braket{{\bf L}}}{\partial t} = -i \braket{\left[{\bf L},{\bf H}\right]},
\end{equation}
where $\left[{\bf L},{\bf H}\right]$ is the commutator of  ${\bf L}$ with ${\bf H}$.

For a closed, isolated physical system the Hamiltonian ${\bf H}$ is time independent (${\bf H}(t) = {\bf H}$), and
the unitary operator is ${\bf U}(t_0, t) = {\bf U}(t - t_0) = e^{-i {\bf H} (t-t_0)}$.
While evolving, a closed system remains in a superposition of states
and preserves their magnitudes and relative angles.

For non-pure states, the evolution of density operators is 
\begin{equation} \label{eq:densityEvolution}
\rho(t) =  {\bf U}^\dagger(t_0, t) \rho(t_0) {\bf U}(t_0, t).
\end{equation}

In most physical as well as in audio applications we have that the system under consideration is driven by external forces, such as a changing magnetic field or a vocal gestural articulation. In such cases of closed non-isolated systems~\cite{breuer}, the Hamiltonian ${\bf H}$ is time dependent. 
The states change under the effect of the external forces, which determine
the change of probabilities, and the Hamiltonian controls the
evolution process.

With a commutative Hamiltonian ($\left[{\bf H}(0),{\bf H}(t)\right] =
0 $), the time evolution can be expressed as 
\begin{equation} \label{stateEvolution2}
  \ket{\psi(t)} = e^{-i\int_0^t {\bf H}(\tau)d\tau}\ket{\psi(0)} = {\bf U}(0, t) \ket{\psi(0)}.
\end{equation}

In general, if the operators $\bf A$ and $\bf B$ do not commute (i.e., $\left[{\bf A},{\bf B}\right] \neq 0$) we have that
$e^{\bf A} e^{\bf B} \neq e^{{\bf A}+{\bf B}}$. Since the evolution
between two time points $0$ and $t$ can be split at an intermediate
time $t^*$, if $e^{-i\int_0^t {\bf H}(\tau)d\tau} = e^{-i\int_0^{t^*}
  {\bf H}(\tau)d\tau -i\int_{t^*}^t {\bf H}(\tau)d\tau} \neq
e^{-i\int_0^{t^*} {\bf H}(\tau)d\tau} e^{ -i\int_{t^*}^t {\bf
    H}(\tau)d\tau}$ then it means that an explicit solution in terms
of an integral can not be found.  Our approach is to consider time
segments where the Hamiltonian is locally commutative, and to compute
the time evolution segment by segment in terms of an integral.

\subsubsection{Phon in utterance field}
Similarly to a spin in a magnetic field, when a phon is part of an
utterance, it has an energy that depends on its orientation. We can
think about it as if it was subject to restoring forces, and its
quantum Hamiltonian is
\begin{equation}\label{potentialEnergy}
  {\bf H} \propto \overline{\sigma} \cdot \overline{B} =  \sigma_x B_x + \sigma_y B_y + \sigma_z B_z ,
\end{equation}
where the components of the field $\overline{B}$ are named in analogy with the magnetic field.

Consider the case of potential energy only along $z$:
\begin{equation}
 {\bf H} = \frac{\omega}{2} \sigma_z. 
\end{equation}
To find how the expectation value of the phon varies in time, we expand the observable ${\bf L}$ in~(\ref{eq:Lcomm}) in its components to get
\begin{align}
  \braket{\dot{\sigma}_x}&=-i\braket{\left[\sigma_x,{\bf H}\right]}=-\omega \braket{\sigma_y} \\
  \braket{\dot{\sigma}_y}&=-i\braket{\left[\sigma_y,{\bf H}\right]}=\omega \braket{\sigma_x} \nonumber \\ 
  \braket{\dot{\sigma}_z}&=-i\braket{\left[\sigma_z,{\bf H}\right]}= 0, \nonumber
\end{align}
which means that the expectation values of $\sigma_x$ and $\sigma_y$
are subject to temporal precession around $z$ at angular velocity
$\omega$.  In phon terms, the expectation
value of $\sigma_z$ steadily keeps the pitch if there is no potential
energy along turbulence and myoelastic pulsation.

A potential energy along all three axes can be expressed as
\begin{equation}\label{hamiltonianMagnetic}
  {\bf H} = \frac{\omega}{2}  \overline{\sigma} \cdot \overline{n} = \frac{\omega}{2} \begin{bmatrix} n_z &  n_x - i n_y \\ n_x + i n_y & -n_z \end{bmatrix},
\end{equation}
whose energy eigenvalues are $E_j = \pm \frac{\omega}{2}$, with energy eigenvectors $\ket{E_j}$.

An initial state vector (phon) $\ket{\psi(0)}$ can be expanded in the energy eigenvectors as
\begin{equation}
  \ket{\psi(0)} = \sum_j \alpha_j(0) \ket{E_j},
\end{equation}
where $\alpha_j(0) = \braket{E_j|\psi(0)}$, and the time evolution of state turns out to be
\begin{equation} \label{stateEvolution1}
  \ket{\psi(t)} = \sum_j \alpha_j(t) \ket{E_j} = \sum_j \alpha_j(0) e^{-iE_jt}\ket{E_j}.
\end{equation}

\subsection{Measurement}
Given that time evolution of states is governed by the unitary
transformation~(\ref{unitaryTransformation}) and by the Schr\"odinger
equation~(\ref{eq:timedepSchr}), the measurement postulate of quantum
mechanics~\cite{cariolaro2015quantum} states that a measurement is
represented by an operator (a projector) that acts on the state and
that causes its collapse onto one of its eigenvectors. 
  
A projector system $\Pi_i$ in the (Hilbert) space of
states is Hermitian, idempotent, and complete.
If the system is in state $\ket{\psi}$ before measurement, the
probability that the outcome of a measurement through a projector
system returns $j$ is
\begin{equation}
  p_m(j|\psi) = 
  \bra{\psi} \Pi_j \ket{\psi}
\end{equation}
and, as a result of the measurement, the
system collapses in state $\psi^{(j)}_{post} = \frac{\Pi_j \ket{\psi}
}{\sqrt{p_m(j|\psi)}}$.

Given an orthonormal basis of measurement vectors $\ket{a_j}$, the
elementary projectors are $\Pi_j = \ket{a_j} \bra{a_j} $, $p_m(j|\psi)
= |\braket{\psi | a_j}|^2 $, and the system (by neglecting a unitary
phasor) collapses into $\psi^{(j)}_{post} = \ket{a_j}$.

If the system is in a pure state
\begin{equation}
  p_m(j|\psi) = \braket{\psi \mid
    \Pi_j \mid \psi} = Tr[\rho \Pi_j].
\end{equation}
If the system is in a mixed
state, the outcome of measurement is formulated as a random variable
conditioned by a given state:
\begin{equation}
  p_m(j|\psi_k) = \braket{\psi_k \mid \Pi_j \mid \psi_k} = \\
  Tr[\ket{\psi_k} \bra{\psi_k} \Pi_j]
\end{equation}
and by averaging over all
components of the mixed state we get 
\begin{equation}\label{probabilityMeasuringMixed}
  p_m(j|\rho) = \sum _k p_k p_m(j|\psi_k) =  Tr[\rho \Pi_j].
\end{equation}
If the outcome of measurement is $j$, the system collapses into the new ensemble of states
represented by the density operator
\begin{equation}\label{densityCollapsed}
\rho^{(j)}_{post} = \frac{\Pi_j \rho \Pi_j}{Tr[\rho  \Pi_j] } .   
\end{equation}

\subsection{Audio measurement and evolution}
The mathematics of quantum mechanics can be used to describe and develop some operations of audio signal processing, aimed at segregating components or streams from raw audio. The concepts of quantum measurement and temporal evolution of quantum states can be recast in audio and phonetic terms if we can rely on an audio analysis/synthesis system that permits the extraction and manipulation of slowly varying features such as pitch salience or spectral energy.

\subsubsection{Non-commutativity and autostates}\label{sec:app}

\begin{figure}
 \centerline{{
\includegraphics[width=0.6\columnwidth]{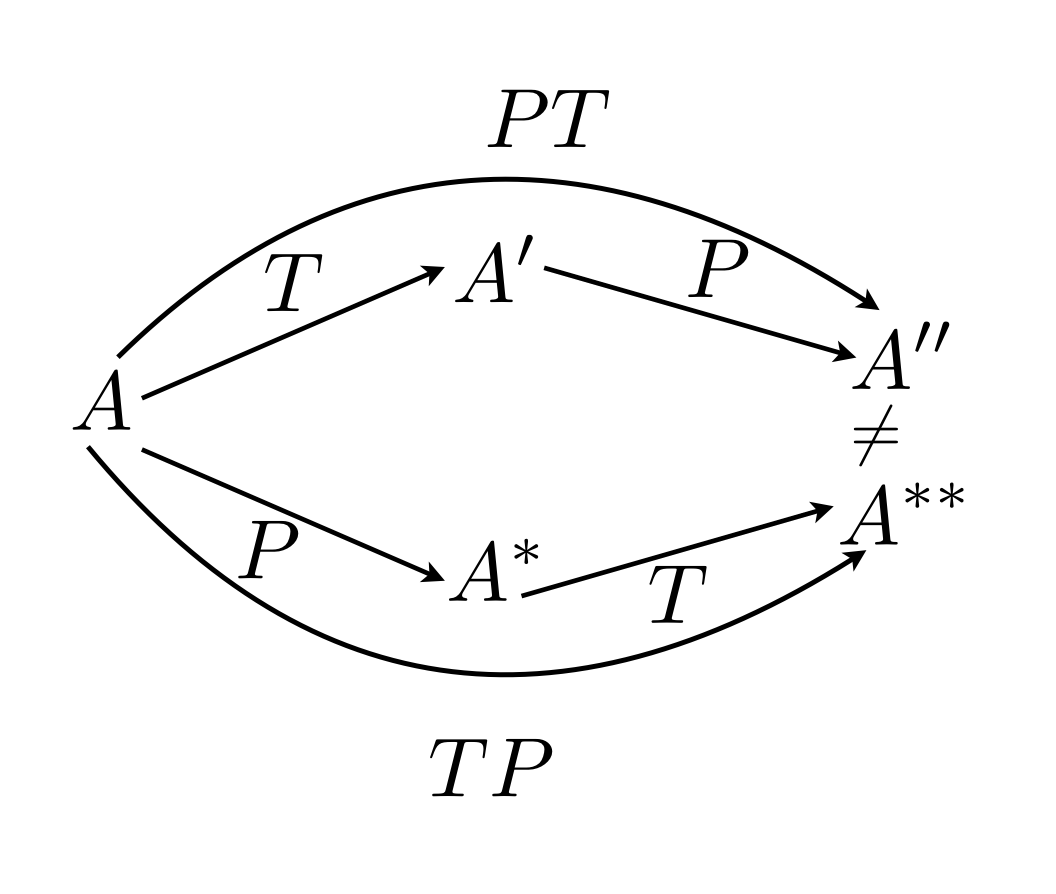}}}
 \caption{A non-commutative diagram representing the non-commutativity of measurements of phonation ($P$) and turbulence ($T$) on audio $A$.}
 \label{commutator1}
\end{figure}

We expect that measurement operators along different axes do not commute: this is the case, for example, of measurements of phonation and turbulence. Let $A$ be an audio \Delete{sample }{segment}. The measurement (by extraction) of turbulence by the operator $T$ leads to $T(A)=A'$. A successive measurement of phonation by the operator $P$ gives $P(A')=A''$, thus $P(A')=PT(A)=A''$. If we perform the measurements in the opposite order, with phonation first and turbulence later, we obtain $TP(A)=T(A^{\ast})=A^{\ast\ast}$. We expect that $[T,P]\neq0$, and thus, that $A^{\ast\ast}\neq A''$. The diagram in figure~\ref{commutator1} shows non-commutativity in the style of category theory.

Besides the compact diagrammatic representation, we can  describe such a non-commutativity in terms of projectors $\Pi_T,\,\Pi_P$:
\begin{equation}\label{commutator_projector}
\begin{split}
&\Pi_T\left( \Pi_P\ket{A} \right) = \ket{T}\braket{T|P}\braket{P|A} = \braket{T|P}\ket{T}\braket{P|A}\neq\\
& \Pi_P\left( \Pi_T\ket{A} \right) = \ket{P}\braket{P|T}\braket{T|A}=\braket{P|T}\ket{P}\braket{T|A}.
\end{split}
\end{equation}
{Given that $\braket{T|P}$ is a scalar and $\braket{P|T}$ is its complex conjugate, and that $\ket{P}\bra{T}$ is generally non-Hermitian, we get
\begin{equation}
\begin{split}
[\Pi_T,\Pi_P] = \ket{T}\braket{T|P}\bra{P} - \ket{P}\braket{P|T}\ket{T} = \\
 =  \braket{T|P}\ket{T} \bra{P} - \braket{P|T} \ket{P}\bra{T} \neq 0.
\end{split}
\end{equation}
}

\begin{figure}
 \centerline{{
\includegraphics[width=1.1\columnwidth]{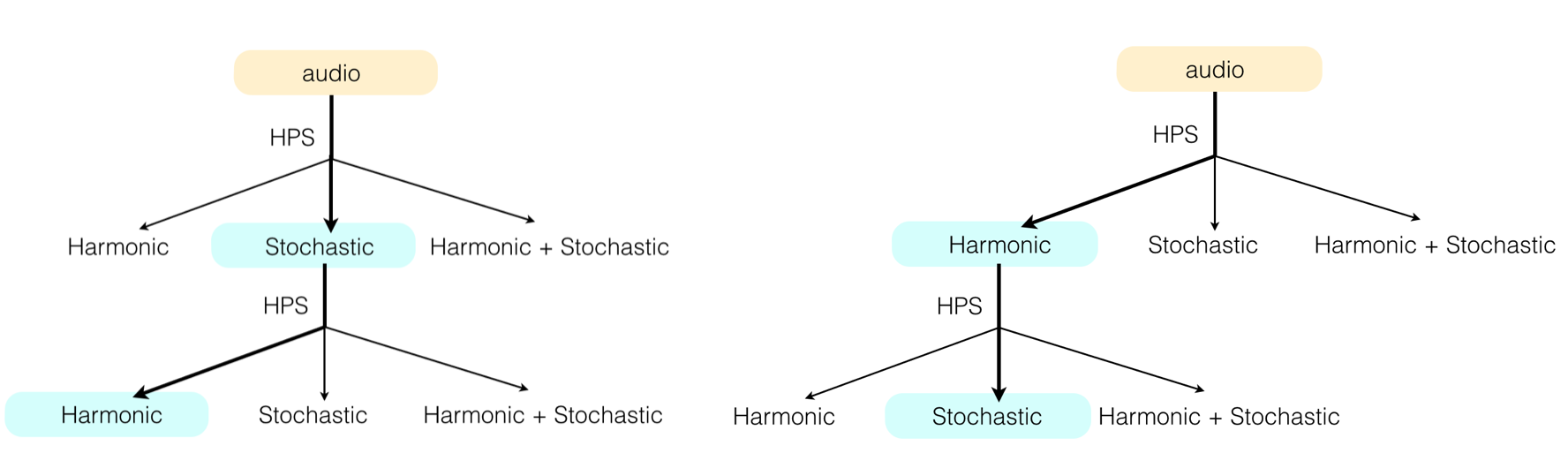}}}
 \caption{On the left, an audio segment is analyzed via the HPS model. Then, the stochastic part is submitted to a new analysis. In this way, a measurement of phonation follows a measurement of turbulence. On the right, the measurement of turbulence follows a measurement of phonation. {This can be described via projectors through equation~(\ref{commutator_projector}), and diagrammatically in fig.~\ref{commutator1}.}}
 \label{commutator1b}
\end{figure}

\begin{figure}
 \centerline{{
\includegraphics[width=\columnwidth]{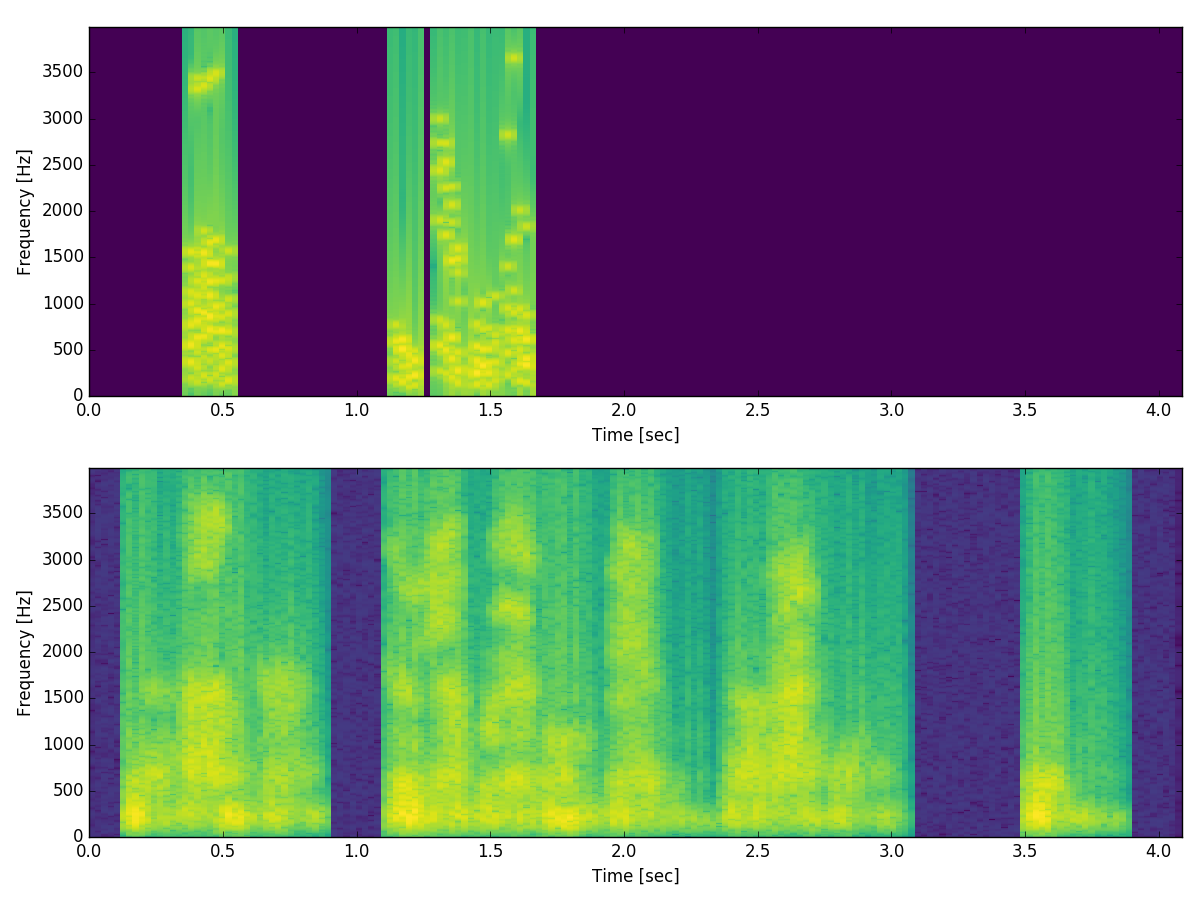}}}
 \caption{On the top, the spectrogram corresponding to a measurement of phonation $P$ following a measurement of turbulence $T$, leading to $PT(A)=A''$.
 On the bottom, the spectrogram corresponding to a measurement of turbulence $T$ following a measurement of phonation $P$, leading to $TP(A)=A^{\ast\ast}$. }
 \label{commutator2}
\end{figure}

Measurements of phonation and turbulence can be actually performed using the sines + noise (a.k.a., Harmonic Plus Stochastic -- HPS) model~\cite{bonada2011spectral}. The order of operations is visually described in figure~\ref{commutator1b}.
The measurement of phonation is performed through the extraction of the harmonic component in the HPS model, while the measurement of turbulence is performed through the extraction of the stochastic component with the same model.
The spectrograms for $A''$ and $A^{\ast\ast}$ in Figure \ref{commutator2} show the results of such two sequences of analyses on a segment of female speech,\footnote{\url{https://freesound.org/s/317745/}. \newline Hann window of 2048 samples, FFT of 4096 samples, hop size of 1024 samples.} confirming that the commutator $\left[T,P\right]$ is non-zero.

\begin{figure}
 \centerline{{
\includegraphics[width=\columnwidth]{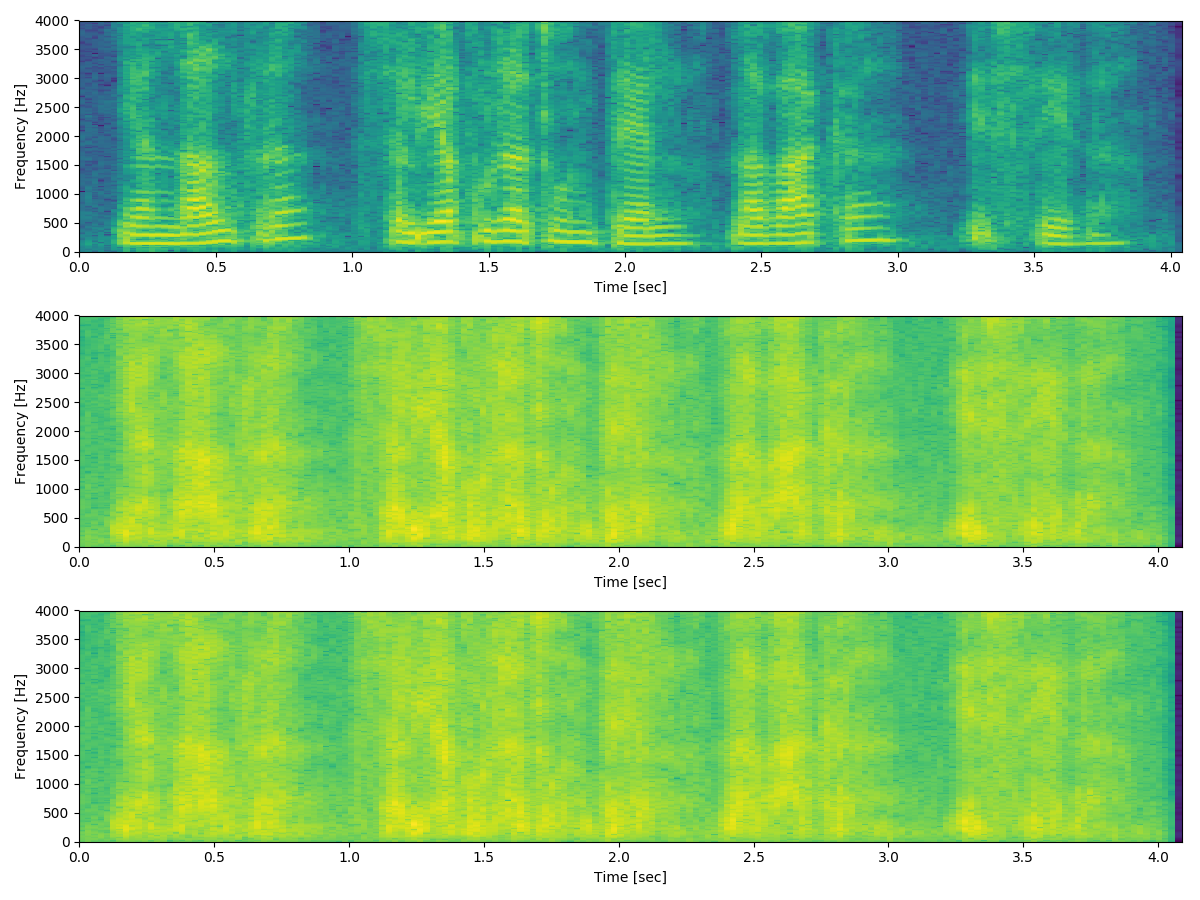}}}
 \caption{Top: spectrum of the original sound signal (a female speech), Center: the stochastic component, derived from harmonic plus stochastic analysis (HPS), as the effect of a destructive measure, and Bottom: the stochastic component of the stochastic component itself. The last two spectra are very close.}
 \label{spectra}
\end{figure}

Essentially, if we adopt the HPS model and skip the final step of addition and inverse transformation, we are left with something that is conceptually equivalent to a quantum destructive measure.
Let $St$ be the filter that extracts the stochastic part from a signal. As figure~\ref{spectra} shows, the spectrogram of $St(x)$ is visibly different from the spectrogram of $x$. Conversely, if we apply $St$ once more, we get a spectrum that does not change much: $St^2(x)=St(St(x))\sim St(x)$. If we transform back from the second and third spectrograms of figure~\ref{spectra}, we get sounds that are very close to each other.  In fact, ideally, $St^2(x)=St(x)$. It means that, after a measure of the non-harmonic component of some signal, the output-signal can be considered as an autostate, and it confirms that the projection operator is idempotent. If we perform the measure again and again, we still get the same result. Such a measure operation provokes the collapse of a hypothetical underlying wave function, which is originally a superposition of states, and is reduced to a single state upon measurement. The importance of the autostates in this framework is connected with the concept of quantum measures, which may become practically feasible through a set of audio-signal analysis tools.

\subsubsection{Hamiltonian streaming}\label{sec:hamiltonian_streaming}
{Let us consider a quantum state vector $\ket{\psi(t)}$, that evolves in time according to the Schr\"{o}dinger equation~(\ref{eq:timedepSchr}). The time evolution can be represented by the unitary operator ${\bf U}(t_0, t)$ of equation~(\ref{unitaryTransformation}).

If we choose a particular, commutative Hamiltonian, the time evolution can be expressed by an integral, as in equation~(\ref{stateEvolution2}). A time-independent Hamiltonian such as the one leading to~(\ref{stateEvolution1}) would not be very useful, both because forces indeed change continuously and because this would lead to oscillatory solution. Similarly to what has been done by Youssry et al.~\cite{youssry}, the Hamiltonian can be chosen to be time-dependent yet commutative (i.e., $\left[ {\bf H}(0), {\bf H}(t) \right] = {\bf H}(0) {\bf H}(t) - {\bf H}(t) {\bf H}(0) = 0$), so that a closed-form solution to state evolution can be obtained. 
A simple choice is that of a Hamiltonian such as
\begin{equation}\label{Hamiltonian_g_S}
  H(t) = g(t) {\bf S},
\end{equation}
with ${\bf S}$ a time-independent Hermitian matrix.  A function $g(t)$ that ensures convergence of the integral in~(\ref{stateEvolution2}) is the damping
\begin{equation} \label{damping}
  g(t) = e^{-t}.
\end{equation}
In an audio application, we can consider a slice of time and the initial and final states for that slice. We should look for a Hamiltonian that leads to the evolution of the initial state into the final state. In image segmentation~\cite{youssry}, where time is used to let each pixel evolve to a final foreground-background assignment, the Hamiltonian is chosen to be
\begin{equation}
  H = e^{-t} f({\bf x})  \begin{bmatrix} 0 &  -i \\ i & 0 \end{bmatrix}, 
\end{equation}
and $f(\cdot)$ is a two-valued function of a feature vector ${\bf x}$ that contains information about a neighborhood of the pixel. Such function is learned from an example image with a given ground truth.
In audio we may do something similar and learn from examples of transformations: phonation to phonation, with or without pitch crossing; phonation to turbulence; phonation to myoelastic, etc. We may also add a coefficient to the exponent in~(\ref{damping}), to govern the rapidity of transformation. As opposed to image processing, time is the playground of audio processing, and a range of possibilities is open to experimentation in Hamiltonian streaming and audio processing.

The matrix ${\bf S}$ can be set to assume the structure~(\ref{hamiltonianMagnetic}), and  the components of potential energy found in an utterance field can be extracted as audio features. For example, pitch salience can be extracted from time-frequency analysis~\cite{SalamonGomez2012} and used as $n_z$ component for the Hamiltonian. Figure~\ref{twoMostSalient}
shows the two most salient pitches, automatically extracted from a mixture of male and female voice\footnote{\url{https://freesound.org/s/431595/}} using the Essentia library~\cite{bogdanov2013essentia}. Frequent up-down jumps are evident, and they make difficult to track a single voice. Quantum measurement induces state collapse to $\ket{u}$ or $\ket{d}$ and, from that state, evolution can be governed by~(\ref{stateEvolution2}). In this way, it should be possible to mimic human figure-ground attention~\cite{bregman1994auditory,bigand2000divided}, and follow each individual voice, or sound stream.

\begin{figure}
 \centerline{{
\includegraphics[width=\columnwidth]{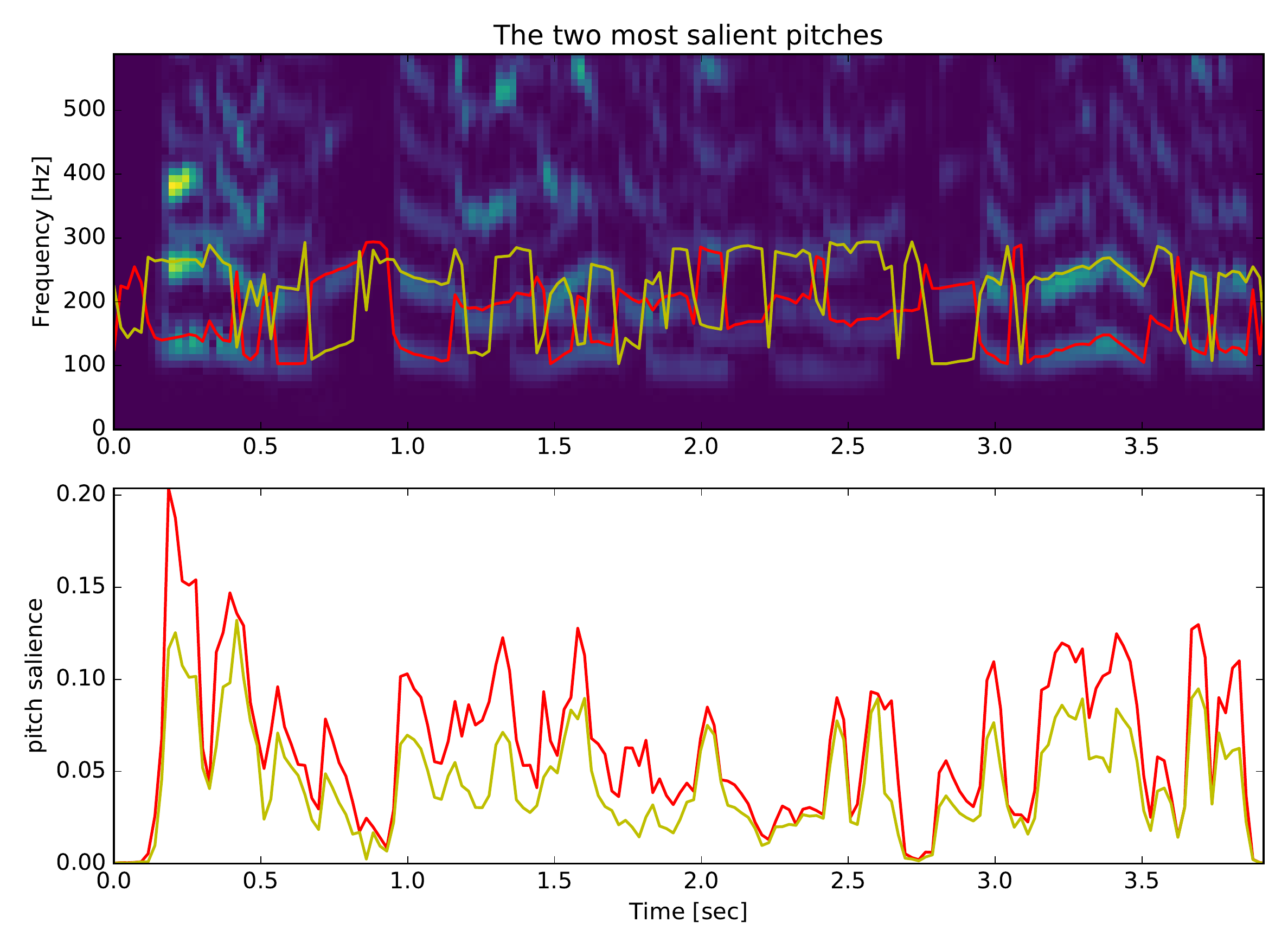}}}
 \caption{Extraction of the two most salient pitches from a mixture of a male voice and a female voice}
 \label{twoMostSalient} 
\end{figure}

\section{Examples}
\label{examples}
This section is intended to illustrate the potential of the Quantum Vocal Theory of Sound in auditory scene analysis and audio effects.\footnote{The reported examples are available, as a {\tt jupyter notebook} containing the full code, on  \href{https://github.com/d-rocchesso/QVTS}{https://github.com/d-rocchesso/QVTS}} 
\subsection{Two crossing glides interrupted by noise}
\label{evolutionPureStates}
In auditory scene analysis, insight into auditory organization is often gained through investigation of continuity effects~\cite{bregman1994auditory}. One interesting case is that of gliding tones interrupted by a burst of noise~\cite{ciocca1987}. Under certain conditions of temporal extension and intensity of the noise burst, a single frequency-varying auditory object is often perceived as crossing the interruption. Specific stimuli can be composed that make bouncing or crossing equally possible, to investigate which between the Gestalt  principles of proximity and good continuity actually prevails. V-shape trajectories (bouncing) are often found to prevail on crossing trajectories when the frequencies at the ends of the interruption match.

To investigate how Hamiltonian evolution may be tuned to recreate some continuity effects, consider two gliding sinewaves that are interrupted by a band of noise.

Figure~\ref{pitchSalienceFunction} (top) shows the spectrogram of such noise-interrupted crossing {\it glissandos}, overlaid with the traces of the two most salient pitches, computed by means of the Essentia library~\cite{bogdanov2013essentia}. Figure~\ref{pitchSalienceFunction} also displays (middle) the computed salience for the two most salient pitches and (bottom) the energy traces for two bands of noise
($1\rm{kHz}$ -- $2\rm{kHz}$, and $2\rm{kHz}$ -- $6\rm{kHz}$).

\begin{figure}
 \centerline{
\includegraphics[width=\columnwidth]{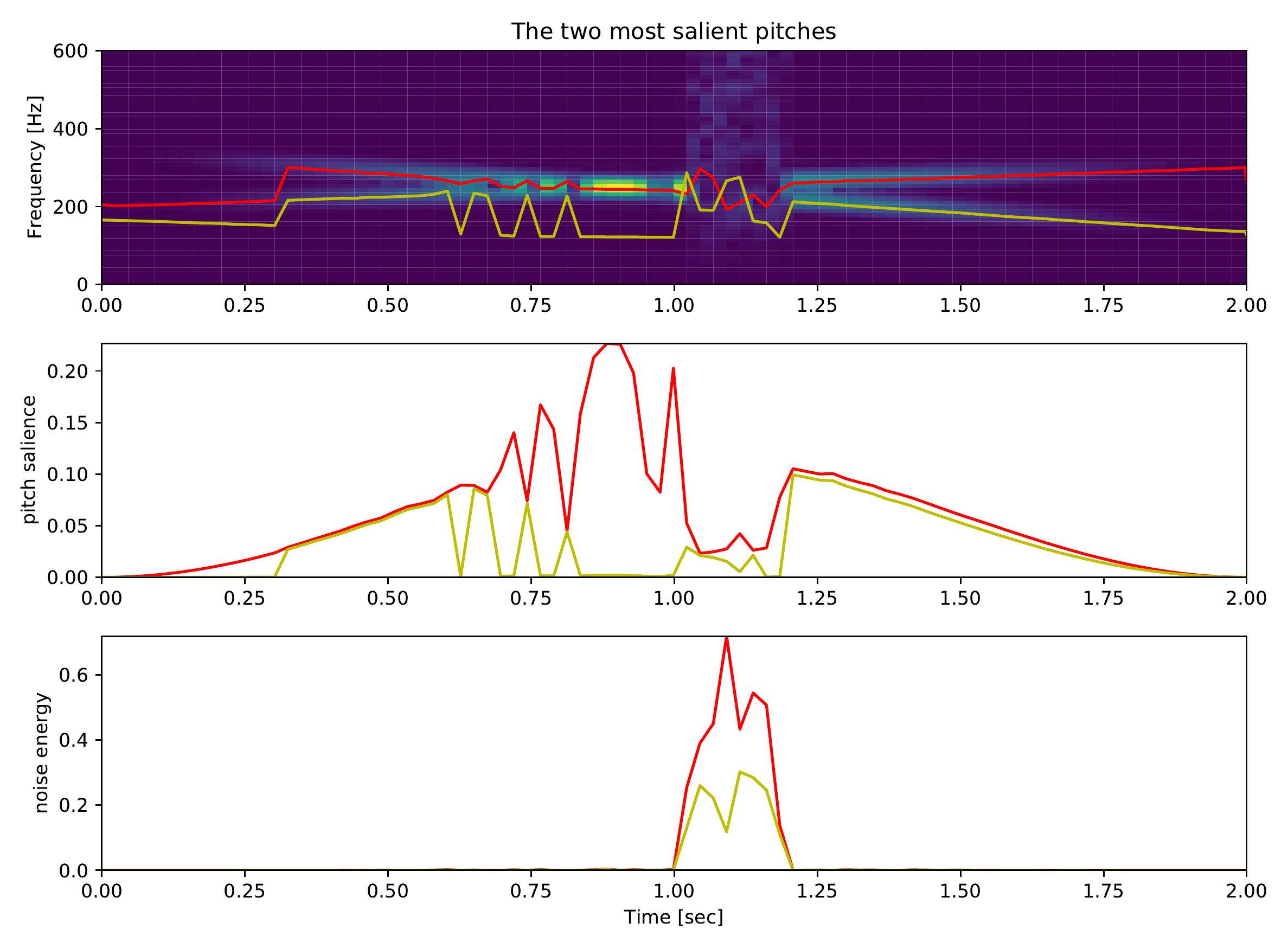}}
 \caption{Tracing the two most salient pitches and noise energy for two crossing glides interrupted by noise}
 \label{pitchSalienceFunction} 
\end{figure}

The elements of the ${\bf S}$ matrix of the Hamiltonian~(\ref{hamiltonianMagnetic}) can be computed (in {\tt Python}) from decimated audio features as
\begin{lstlisting}[language=Python]
decimation = 10 # PARAMETER (hold on)
n_x = energyNoise1[::decimation]
n_x = np.repeat(n_x, decimation)[0:size(energyNoise1)] # turbulence potential
n_y = np.zeros(n_x.size) # n_y is not at play 
n_z = totalSaliences1[::decimation] 
n_z = np.repeat(n_z, decimation)[0:size(energyNoise1)] # pitch potential
\end{lstlisting}
and the time-varying Hamiltonian can be multiplied by a decreasing
exponential $g(m) = e^{-km}$, where $m$ is the frame number, extending
over $M$ frames:
\begin{lstlisting}[language=Python]
k = 0.1 # PARAMETER (damping)
et = exp(-k*np.arange(0,decimation))
et = np.tile(et,int(size(n_z)/decimation))
ett = np.pad(et,(0,size(n_z)-size(et)),'constant',constant_values=(0))
nzc = n_z*ett
nxc = n_x*ett
nyc = n_y*ett
\end{lstlisting}
The resulting turbulence and phonation potentials are depicted in figure~\ref{tbp}.

\begin{figure}
 \centerline{
\includegraphics[width=\columnwidth]{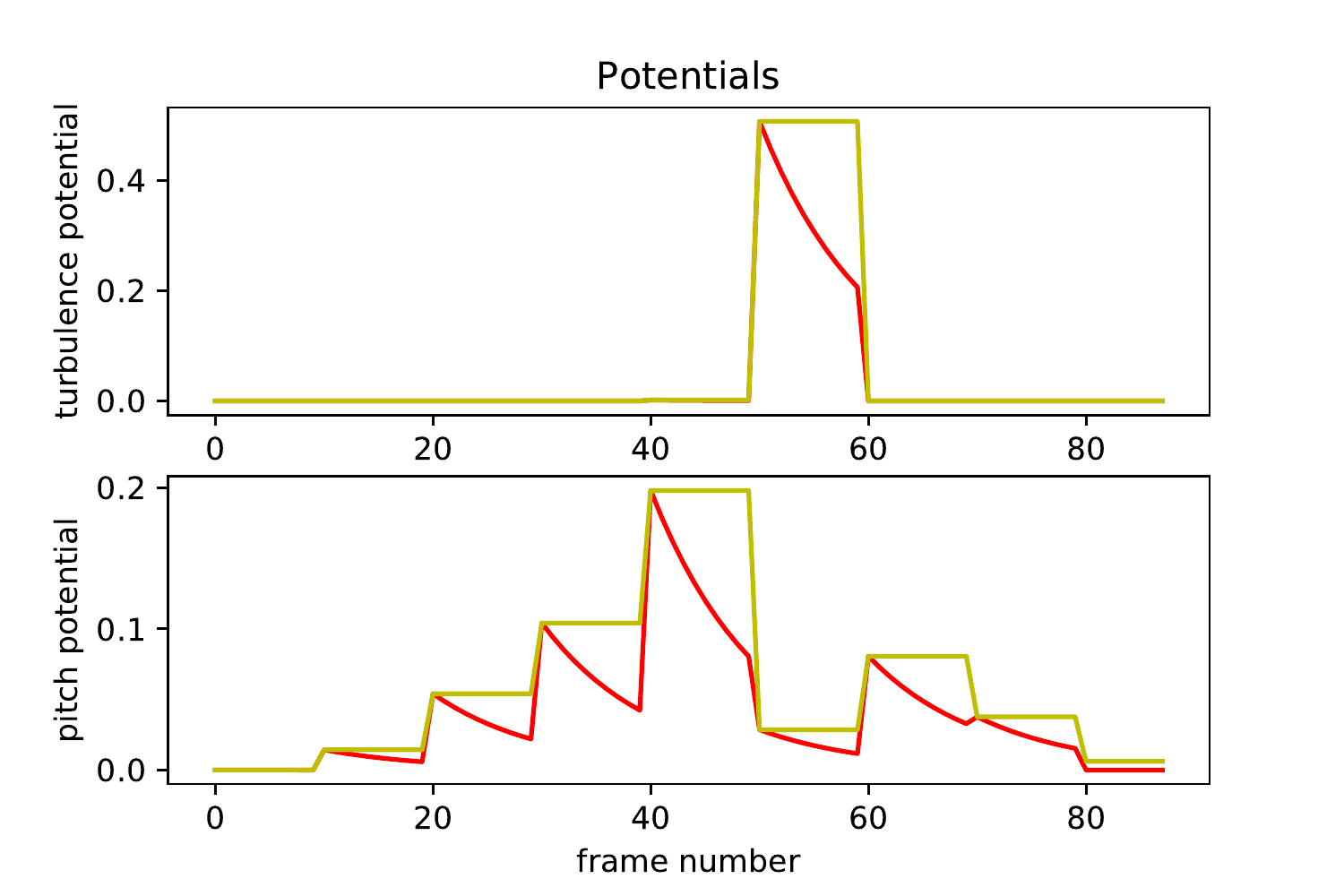}}
 \caption{Potentials of turbulence (top) and phonation (bottom) as functions of frame number}
 \label{tbp} 
\end{figure}

The Hamiltonian time evolution of equation~(\ref{stateEvolution2}) can
be computed by approximating the integral with a cumulative sum:
\begin{lstlisting}[language=Python]
H = np.array([[nzc, nxc - 1j*nyc], [nxc + 1j*nyc, -nzc]]) 
intHc = np.cumsum(H, axis=2)
Uc = intHc
for t in range(size(n_z)):
    Uc[:,:,t] = linalg.expm(-(1j)*intHc[:,:,t]) 
\end{lstlisting}
Choosing an initial state (e.g., pitch-up), the state evolution can be converted into a pitch (phonation) stream, which switches to noise (turbulence) when it goes below a given threshold of pitchiness:
\begin{lstlisting}[language=Python]
threshold = 0.7 # PARAMETER (pitchiness)
hopCollapse = 5 # PARAMETER (decimation of collapses)
sigma_z = [[1, 0], [0, -1]]
sigma_x = [[0, 1], [1, 0]]
for t in range(size(totalFreqs1)): 
    sdiff = norm(transpose(matrix(sT[:, t])) - 
        sigma_z * transpose(matrix(sT[:, t])))
    if (sdiff < threshold or 
        sdiff > (2 - threshold)): 
        # state is pitchy 
        prob = np.square(abs(sT[0,t])) 
        # collapse based on probability
        cstate = np.random.choice(
          np.arange(0, 2), p=[prob, 1-prob]) 
        if (cstate == 0):
            if (t%hopCollapse == 0):
                sT[0,t] = 1.0
                sT[1,t] = 0.0
            <select upper pitch>
        else: 
            if (t%hopCollapse == 0):
                sT[0,t] = 0.0
                sT[1,t] = 1.0
            <select lower pitch>
    else: # state is turbulent
        prob = abs(np.square(np.dot(sT[:,t],
               [1/sqrt(2),1/sqrt(2)]))) 
        cstate = np.random.choice(
          np.arange(0, 2), p=[prob, 1-prob]) 
        if (cstate == 0):
            if (t%hopCollapse == 0):
                sT[0,t] = 1/sqrt(2)
                sT[1,t] = 1/sqrt(2)
            <select lower band of noise>
        else:
             if (t%hopCollapse == 0):
                sT[0,t] = 1/sqrt(2)
                sT[1,t] = -1/sqrt(2)
           <select higher band of noise>
\end{lstlisting}
In the proposed implementation, the free parameters are {\tt decimation}, {\tt k}, {\tt thresh\-old} and {\tt hopCollapse}, the latter being a decimation on the measurements that are a accompanied by a state collapse. This small set of parameters allow to produce a variety of temporal behaviors, well beyond what is possible with a rigid quantum-mechanical encoding of the listening process.

One resulting pitch stream evolution from pitch-up is depicted in figure~\ref{evolutionFromPitchUp} and it shows a breaking of continuity with bouncing. A first pitch oscillation is visible around second 0.75 when the two sine waves are beating close to each other, although phonation sticks to pitch-up. Then, when the noise interruption arrives after second 1.00, pitch attribution as well as phonation become uncertain. Such state of pitch confusion persists almost until second 1.40, well beyond the noise interruption, with occasional commutations to a turbulent state. After the noise shock has been forgotten, the tracking process sticks back to pitch-up, thus preferring a bouncing over a crossing trajectory. Occasionally, due to the inherent randomness of the process, the crossing trajectory may be chosen by the tracking process. The relative probability of bouncing versus crossing depends both on the characteristics of the stimulus (slopes of sinusoidal trajectories, width of the noise break, relative amplitude between noise and sines) and on some model parameters such as the relaxation coefficient {\tt k} of the exponential and the probability {\tt threshold} for collapsing the measure to phonation rather than turbulence.

This example, and some other experiments run with different parameters, show that the quantum-vocal model can reproduce some relevant phenomena of auditory continuity~(\cite{warren2008}, ch. 6), which are attributable to neural reallocation. The confusion between phonation and turbulence that extends well beyond the interruption is consistent with the known perceptual fact that bursts of noise are not  precisely located as referred to a tonal transition, with errors up to a few hundred milliseconds~\cite{vicario1963}. 

\begin{figure}
 \centerline{
   \includegraphics[width=\columnwidth]{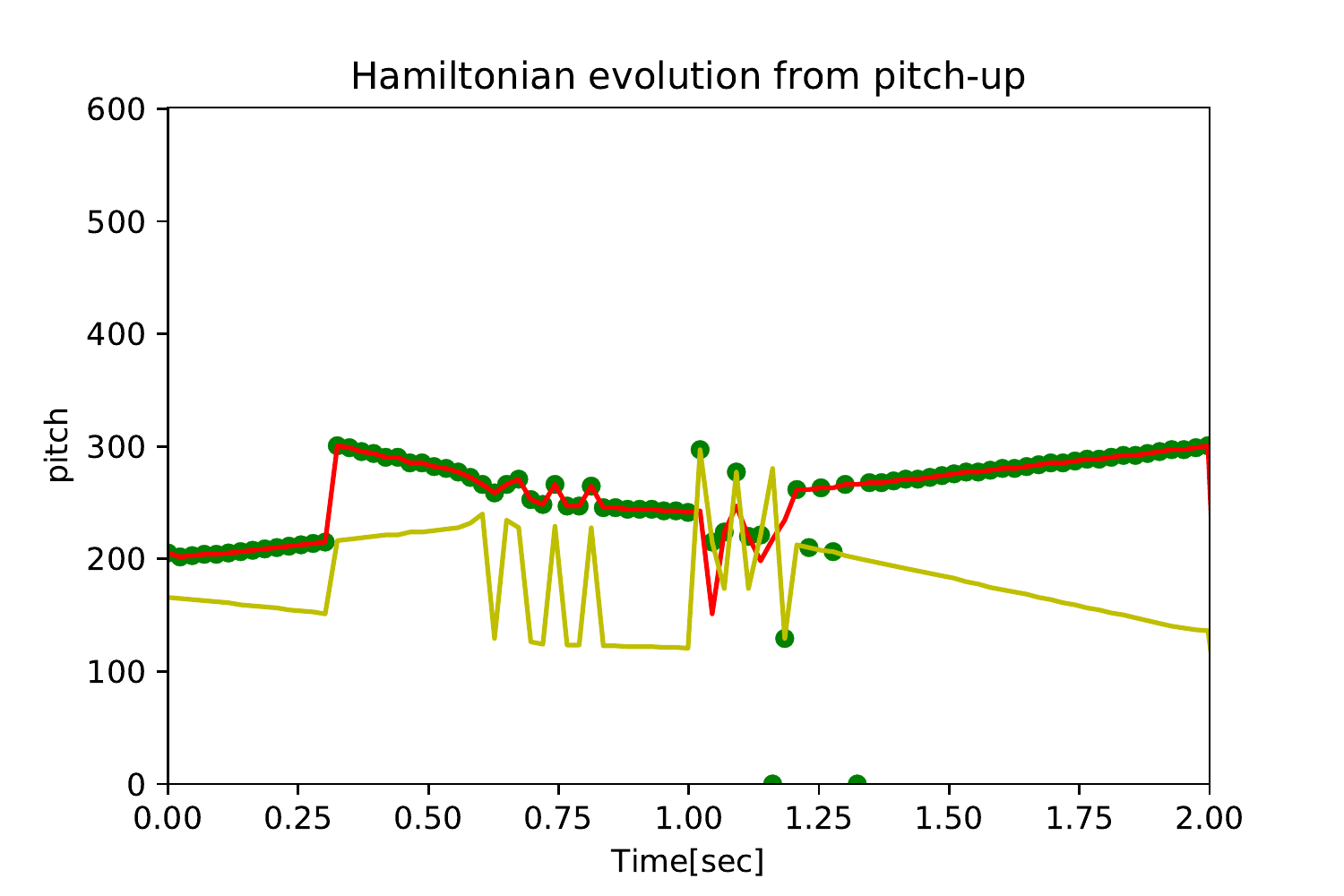}}
 \caption{Tracking the phonation state under Hamiltonian evolution from pitch-up}
 \label{evolutionFromPitchUp} 
\end{figure}}

\subsection{Mixed as in a mixer}
Given an audio scene such as that of the two crossing glides interrupted by noise (figure~\ref{pitchSalienceFunction}), we may follow the Hamiltonian evolution from an initial state that is known only probabilistically. For example, at time zero we may start from a mixture of $\frac{1}{2}$ pitch-up and $\frac{2}{3}$ pitch-down. The density matrix~(\ref{exampleRho}) would evolve according to equation~(\ref{eq:densityEvolution}), where the unitary operator ${\bf U}(0,t)$ is defined as in~(\ref{stateEvolution2}).
When a pitch measurement is taken, the outcome would be up or down according to equation~(\ref{probabilityMeasuringMixed}), and the density matrix that results from collapsing would be given by equation~(\ref{densityCollapsed}).

\begin{figure}
 \centerline{\includegraphics[width=\columnwidth]{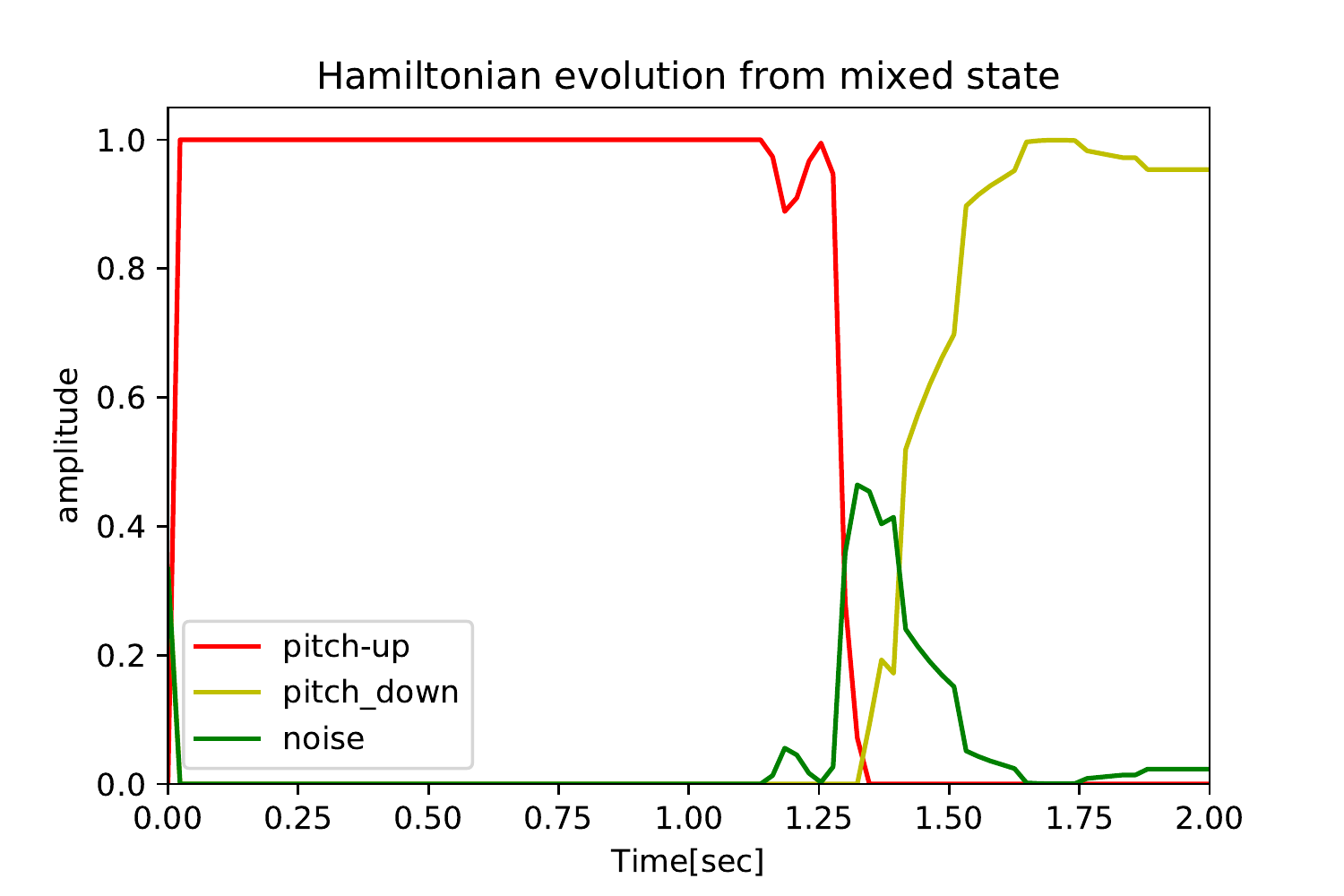}}
 \caption{Amplitudes of components pitch-up, pitch-down, and noise resulting from a Hamiltonian evolution from a mixed state}
 \label{evolutionMixed} 
\end{figure}

\begin{figure}
 \centerline{\includegraphics[width=\columnwidth]{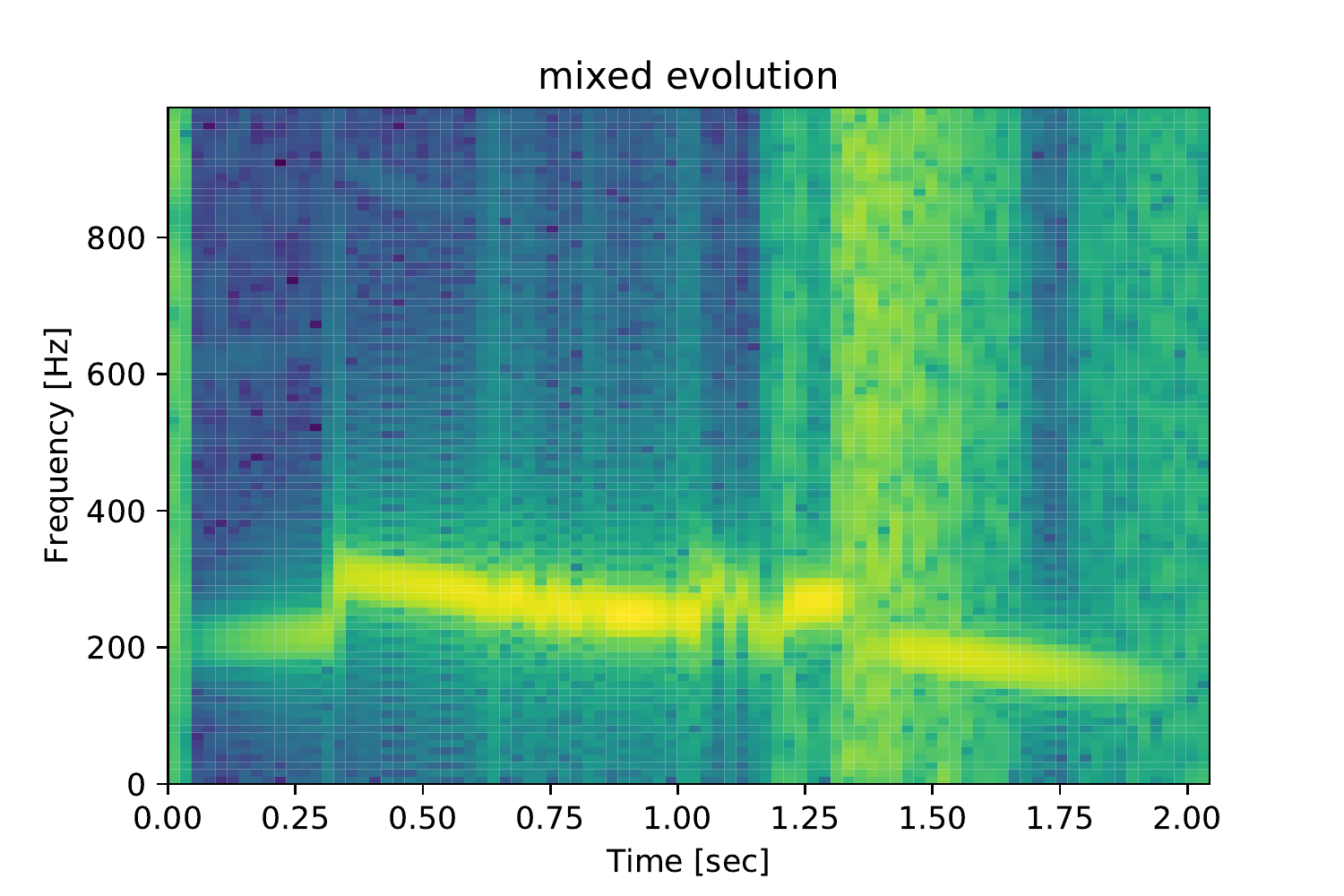}}
 \caption{Spectrogram of the sonification of the Hamiltonian evolution from a mixed state, using the component amplitudes depicted in figure~\ref{evolutionMixed}}
 \label{evolutionMixedSpecgram} 
\end{figure}

The density matrix can be made audible in various ways, thus sonifying the Hamiltonian evolution. For example, the completely chaotic mixed state, corresponding to the half-identity matrix $\rho = \frac{1}{2} {\bf I}$, can be made to sound as noise, and the pure states can be made to sound as the upper or the lower of the most salient pitches. These three components can be mixed for intermediate states. If $p_u$ and $p_d$ are the respective probabilities of pitch-up and pitch-down as encoded in the mixed state, the resulting mixed sound can be composed by a noise having amplitude $\min{(p_u, p_d)}$, by the upper pitch weighted by $p_u - \min{(p_u, p_d)}$, and by the lower pitch weighted by $p_d - \min{(p_u, p_d)}$.
One example of such evolution from a mixed state, with periodic measurements and collapses that reset the density matrix is depicted in figure~\ref{evolutionMixed}. The analyzed audio scene and the model parameters, including the computed Hamiltonian, are the same as used in the evolution of pure states described in section~\ref{evolutionPureStates}. The depicted instance of evolution, if sonified by controlling the amplitudes of the extracted two most salient pitches and of a noise, results in a prevailing downward tone and in a delayed and slowly-decreasing burst of noise (figure~\ref{evolutionMixedSpecgram}).

\section{Conclusion and perspective}\label{sec:concl}
The components of phonation, turbulence, and supraglottal myoelastic vibrations (and clicks) can be found, in some form and possibly in superposition, in all kinds of vocal sound. Since the voice gives a possibility for an embodied representation of sound in general, we can use the three aforementioned basic phonetic components as general sound descriptors. In this work, we proposed the phon as an analogue of a particle spin, where the phonetic components appear to be aligned along the $x$, $y$, and $z$ spin measurement directions. As such, the phon is subject to the mathematical formalism and to the postulates of quantum mechanics, and it can be used to describe sonic processes. Such description is of higher level and exploits a conventional analysis/synthesis framework based on spectral modeling. In particular, we have shown how a time-varying Hamiltonian, that governs the temporal evolution of auditory streams, can be constructed from features that are extracted from spectral modeling. 

In a computational realization of the quantum-inspired operators and processes, the manipulation of a few parameters allows to extract a variety of components from complex audio scenes. The simple examples that we provided show how some relevant auditory streaming phenomena can be modeled and reproduced, but extensive experimentation is definitely required to verify how useful a Quantum Vocal Theory of Sound could be in auditory scene analysis.  A large range of possibilities is also open to the creative processing of audio materials through the sonification of the extracted streams and events. As compared to analysis/synthesis frameworks based on spectral processing, here we work at a higher level corresponding to fewer descriptors whose evolution and intertwinement are mathematically defined. The statistical nature of measurement, in evolutions of pure or mixed states under time-varying force fields, leads naturally to the synthesis of ensembles of audio processes, all derived and somehow echoing the  original audio material. If we successfully model some auditory phenomena, such as continuity effects or temporal displacement, by temporal phon evolution, and if we render these evolutions back to sound we may somehow say that we listen to possible auditory processes. However, in creative applications we are not bound to mimic auditory processes and we can also depart from quantum orthodoxy in many possible different ways. 

The proposed theory enhances the role of quantum theory and of the underlying mathematics as a connecting tool between different areas of human knowledge. By flipping the wicked problem of finding intuitive interpretations of quantum mechanics, we aimed at using quantum mechanics to interpret something that we have embodied, intuitive knowledge of.

\bibliographystyle{plainnat-modified}
\bibliography{qvtsBiblio}

\end{document}